%% file: bare_jrnlv2.tex
\documentclass[journal]{IEEEtran}
\usepackage{cite}

\ifCLASSINFOpdf
\else
\fi
\usepackage{amsmath}
\usepackage[hidelinks]{hyperref}
\usepackage[utf8]{inputenc}
\usepackage{amssymb}
\usepackage[pdftex]{graphicx}
\usepackage{makecell}
\usepackage[short]{optidef}
\makeatletter
\let\MYcaption\@makecaption
\makeatother

\usepackage[font=footnotesize]{subcaption}

\makeatletter
\let\@makecaption\MYcaption
\makeatother
\usepackage{comment}
\usepackage[table,dvipsnames]{xcolor}
\usepackage[nolist]{acronym}
\usepackage[]{nomencl}
\usepackage{soul} 
\sethlcolor{Apricot} 

\RequirePackage{ifthen}
\renewcommand{\nomgroup}[1]{%
	\ifthenelse{\equal{#1}{S}}{\vspace{8pt} 
		\item[\textit{Sets}]}{%
		\ifthenelse{\equal{#1}{I}}{\vspace{8pt} 
			\item[\textit{Indices}]}{%
			\ifthenelse{\equal{#1}{P}}{\vspace{8pt} \item[\textit{Parameters}]}{%
				\ifthenelse{\equal{#1}{F}}{\vspace{0pt} 
					\item[\textit{Functions}]}{%
					\ifthenelse{\equal{#1}{V}}{\vspace{8pt} \item[\textit{Variables}]}{}}}}}}

\definecolor{bg}{rgb}{0.97,0.97,0.97}
\usepackage{float}
\newfloat{listi}{thp}{lop}
\floatname{listi}{Listing}

\usepackage{pgfplots}
\pgfplotsset{compat=newest}
  \usetikzlibrary{plotmarks}
  \usetikzlibrary{arrows.meta}
  \usepgfplotslibrary{patchplots}

\usepackage{multirow}
\usepackage{multicol}
\usepackage{ctable} 
\newcommand{\q}{\mathsf{q}}
\renewcommand{\d}{\mathsf{d}}
\newcommand{\dq}{\mathsf{dq}}
\newcommand{\dqz}{\mathsf{dq0}}

\begin{document}
%
\title{\texttt{PowerSimulationsDynamics.jl} - An Open Source Modeling Package for Modern Power Systems with Inverter-Based Resources}
%
%
%
\author{Jose~Daniel~Lara,~\IEEEmembership{Senior Member,~IEEE,}
        Rodrigo Henriquez-Auba ~\IEEEmembership{Member,~IEEE,} Matthew Bossart ~\IEEEmembership{Student Member,~IEEE,} Duncan S.~Callaway ~\IEEEmembership{Member~IEEE}, Bri-Mathias Hodge ~\IEEEmembership{Senior Member~IEEE},
        and~Clayton Barrows,~\IEEEmembership{Senior Member,~IEEE,}
\thanks{J.~D.~Lara, R.~Henriquez-Auba, B-M. Hodge and C.~Barrows are with the Grid Planning and Analysis Center at the National Renewable Energy Laboratory.}
\thanks{M. Bossart and B-M. Hodge are with the Department of Electrical, Computer, and Energy Engineering at the University of Colorado Boulder.}
\thanks{D.~S.~Callaway and J.D. Lara are with the Energy and Resources Group at the University of California Berkeley.}
\thanks{D.~S.~Callaway and R.~Henriquez-Auba are with Department of Electrical, Computer Engineering at the University of California Berkeley.}
}

%
%

\markboth{IEEE Transactions in Power Systems}%
{Lara \MakeLowercase{\textit{et al.}}: PowerSimulationsDynamics.jl - An Open Source Modeling Package for Modern Analysis of Power Systems with Inverter-Based Resources}
%



\maketitle

\begin{abstract}
In this paper we present the development of an open-source simulation toolbox, \texttt{PowerSimulationsDynamics.jl}, to study the dynamic response of power systems, focusing on the requirements to model systems with high penetrations of Inverter-Based Resources (IBRs). \texttt{PowerSimulationsDynamics.jl} is implemented in Julia and features a rich library of synchronous generator, inverter, and load models. In addition, it allows the study of quasi-static phasors and electromagnetic $\dq$ models that use a dynamic network representation. Case studies and validation exercises show that \texttt{PowerSimulationsDynamics.jl} results closely match other commercial and open-source simulation tools.
\end{abstract}

\begin{IEEEkeywords}
Inverter-Based Resources, Low-Inertia Power Systems, Software Implementation, Positive Sequence, Electromagnetic Simulations.
\end{IEEEkeywords}

%
\IEEEpeerreviewmaketitle
\input{acros2}
\section{Introduction}
%
%
%
%
\IEEEPARstart{S}imulations have been the primary tool for studying power systems dynamics and stability for decades, as the scale and complexity of interconnected power systems limit the application of analytical techniques. \acf{QSP} and \acf{EMT} are the most commonly used simulation models for power system dynamics studies.
Although there is a wealth of knowledge about models, simulation techniques, and software implementations, continuous advances in computer hardware, programming languages, and numerical integration techniques make the field of power system dynamic simulations a perpetually evolving area. 

With the addition of \acp{IBR} into the electric system, new dynamics in the controls of \acp{IBR} are changing modeling requirements for system-wide stability studies that rely on time-domain simulations~\cite{paolone2020fundamentals, hatziargyriou2020definition}. Furthermore, interactions at higher frequencies, driven by the increasing reliance on \acp{IBR}, underscore the need to revisit simulation methods and models for such systems~\cite{hatziargyriou2020definition}.

Introducing new energy conversion technologies requires further research into power system simulation models and techniques. Professionals in the power system community have shared their observations regarding the limitations of current commercial \ac{QSP} tools for systems featuring significant \ac{IBR} penetrations and weak interconnections. For example, an ERCOT report \cite{ercot_stability} has noted non-convergence and numerical instabilities in grid situations characterized as weak grids. In recent reports about oscillations \cite{oscillations} researchers discuss how models can't adequately explain oscillations for different operating conditions. When \ac{QSP} simulations cannot provide valid results, researchers and practitioners turn to point-on-wave \ac{EMT} models to conduct comprehensive system-wide studies at the cost of significantly increasing the computational complexity, if such large-scale \ac{EMT} simulations are even feasible.  

Power system simulation packages usually include a modeling layer, where the behavior of system elements is specified, and a simulation layer, where the algorithmic solution approach is defined. However, in most cases, the developer has restricted access to the details of these layers, which limits model development to the pre-defined structures of the solution algorithms available. The interdependence between the algorithm and models, software limitations and licensing restrictions highlights the need for a modular and open simulation platform to study high \ac{IBR} penetration power systems. Creating bespoke scripts to solve time domain simulations is a compelling path to explore new models such as the case in \cite{markovic2021understanding}. However, single-use software rarely supports large-scale, reproducible, or expansible experiments beyond their intended use case.

Open-source efforts to implement simulation libraries and packages have been used successfully by the research community to explore new models and algorithms. The Matlab-based library PSAT has pioneered extensible, dynamic modeling for power systems\cite{milano2005psat}. The \ac{QSP} Python-based tool ANDES \cite{cui2020hybrid} offers a modeling approach where a symbolic layer describes the components and a numeric layer performs vector-based numerical computations. Dyna$\omega$o is a hybrid C++/Modelica open-source suite of simulation tools for power systems \cite{Guironnet_2018} that enables \ac{QSP} and $\dqz$-\ac{EMT} simulations. In the Julia ecosystem, \texttt{PowerDynamics.jl} \cite{plietzsch2022powerdynamics} also uses symbolic representation of the dynamic models to provide a two-stage process of symbolic and compiled representation of a dynamic network, similar to the approach used in ANDES. Tools like \texttt{ParaEMT} \cite{paraemt} and \texttt{GridPACK} \cite{gridpack} have focused on the  acceleration of point-on-wave \ac{EMT} and \ac{QSP} respectively through parallelization and the use of High-Performance-Computing. 

However, the level of customizability and modularity of open source tools has some limitations. Most libraries are deeply integrated with the solution methods, akin to commercial software, and it can be very difficult to use them to explore new numerical integration algorithms required to tackle simulations with inverters. In some cases implementing new inverter models requires direct modification of the underlying simulation logic source code. Although there is support for component-level flexibility for generator models, there has been limited focus thus far on inverters. 

\subsection{Scope and Contributions}

In this paper we describe a new modeling platform, \texttt{PowerSimulationsDynamics.jl} (\texttt{PSID.jl}), a Julia-based open-source package designed to exploit support both \ac{QSP} and $\dqz$-\ac{EMT} time-domain simulations of \ac{IBR} dominated systems as defined in \cite{def_paper}. 

The approach developed uses a data model for inverters and generators for component-level customization and flexible \ac{ODE} and Algebraic representations of the network circuit dynamics, exploiting Julia's \emph{multiple dispatch} to support its modeling flexibility. The system model is also flexible, enabling the usage of different numerical integration routines via a common API \texttt{DifferentialEquations.jl} \cite{rackauckas2017differentialequations}. \texttt{PSID.jl} uses automatic differentiation techniques to calculate the Jacobians of existing and custom models. 

This work explains the implementation of \texttt{PSID.jl} and its application to tackle the challenges in modeling systems with high shares of \acp{IBR} at scale. The key contributions to the modeling of state of the art system simulation with large penetration of \ac{IBR} are:
\begin{itemize}
    \item An implementation of \ac{IBR} and machine models based on generic data models that supports developing encapsulated sub-component models. This modular design enables code and model reuse, reducing development requirements and enabling fast and simple prototyping new controls and models and implementation of \acp{UDM}.   
    \item \texttt{PSID.jl} supports model and integration algorithm decoupling that provides residual and mass matrix formulations of the system model, enabling the exploration of solution techniques.
    \item Network circuit model using $\dq$ transformation enables the formulation of \ac{QSP} and balanced \ac{EMT} models for the study of system stability, including fast dynamics and the exploration of stiff model integration algorithms for large models.
\end{itemize}


The rest of the paper is structured as follows: Section \ref{sec:models} discusses the simulation models implemented in \texttt{PSID.jl} including the residuals and mass-matrix formulations. Section \ref{sec:soft} discusses the details of the software design and structure, as well as the modeling details for generators, inverters, loads, and network circuits. The simulation case studies and verification results are shown in Section \ref{sec:results} where we compare the results from \texttt{PSID.jl} with commercial tools to verify the results. Finally, the conclusions are presented in Section \ref{sec:conclusions}. 

\section{Power System Models} \label{sec:models}
The simulation modeling approach in \texttt{PSID.jl} follows from the definitions in \cite{def_paper} where 
the simulation of an interconnected transmission-level system comprised of a collection of balanced 3-phased interconnected components through $RLC$ circuits can be formulated as an initial-value-problem:
\begin{subequations}
\begin{align}
    \frac{d\vec{{x}}}{dt} &= F(\vec{{x}}, \vec{{y}}, \vec{\eta}), \quad \vec{{x}}(t_0) = \vec{{x}}^0, \label{eq:sim1}\\
    \frac{d\vec{{y}}}{dt} &= G(\vec{{x}}, \vec{{y}}, \vec{\psi}), \quad \vec{{y}}(t_0) = \vec{{y}}^0, \label{eq:sim2}
\end{align}
\end{subequations}
\noindent where $\vec{{x}}$ and $F(\cdot)$ represent the device (e.g., \acp{IBR}, machines, loads) states and equations with $\vec{\eta}$ parameters. The circuit states and dynamics of the network are represented as the sub-system $\vec{{y}}$ and $G(\cdot)$ with network parameters $\vec{\psi}$. The general \texttt{PSID.jl} model is a standard current injection model; therefore, $\vec{{y}}$ represents the network voltages and currents. The numerical advantages of current injection models outweigh the complexities of implementing constant power loads for longer-term transient stability analysis and support the modeling of fast network dynamics \cite{milano2010power}.

Given the system model \eqref{eq:sim1}-\eqref{eq:sim2}, a simulation can be defined as follows: given an initial condition for the device and network states $\vec{{x}}(t_0), \vec{{y}}(t_0)$ advance the solution in time $t$ from one point to the next considering a discrete timeline $\{t_0,t_1,\dots,t_n,\dots,T\}$. A simulation requires a stepping algorithm that finds the solution at time $t_{n+1}$ provided the values of the variables at $\{t_k | t_0 \le t_k < t_{n+1}\}$. 

\subsection{Solution methods for time-domain simulations }


The challenges of integrating \acp{IBR} stem from both interactions across time scales such as interactions with transmission circuit electromagnetics  \cite{markovic2021understanding} complicating the study of system dynamics. Most commercial \ac{QSP} tools use Partitioned-Explicit solution methods which are not A-stable \cite{ascher1998computer}. \texttt{PSID.jl} implements a formulation suitable for the exploration of simultaneous solution methods to address the integration challenges of stiff multi-rate systems\footnote{A more detailed discussion on the implications of simulation models for the integration technique can be found in \cite{def_paper}}. This design choice is geared towards the exploration of implicit solution methods that can handle unknown stiffness \textit{a priori} and maintain a workflow similar to the one already used by \ac{QSP} tools. Furthermore, the potential challenge of stiffness requires different approaches depending on the properties of the system being simulated -- for example, stiffness arising from eigenvalues that exhibit large rations negative parts versus cases with large imaginary eigenvalues. The need to provide flexible formulations for power systems simulation models is described in detail in \cite{7386712}, where the author notes the flexibility and algorithmic improvements afforded by a semi-implicit formulation of the dynamic model. \texttt{PSID.jl} is able to formulate \ac{DAE} models of dynamical equations in two different ways which adapt to the requirements of implicit numerical methods using error control and variable step properties. 

\textbullet~{\textbf{Residual Model:}\footnote{In the differential equations literature, this model is also named \emph{semi-explicit} system}. This model is implemented for methods that find the solution to $\text{H}(t, z_t, \frac{d z_t}{d t}) = 0$ at each time step $t$. This formulation distinguishes between a subset of differential states $z_d$ and algebraic states $z_a$, where the differential states are described by at least one derivative, resulting in the following system formulation:
\begin{subequations}
\begin{align}
  \vec{r}_{x_{d}} &= \frac{d\vec{x}_d}{dt} - F_d(\vec{x}_d, \vec{x}_{a}, \vec{y}_d, \vec{y}_a, \vec{\eta}) \label{eq:sim_res1}\\
  \vec{r}_{x_{a}}  &=  F_a(\vec{x}_d, \vec{x}_a, \vec{y}_d, \vec{y}_a, \vec{\eta}) \label{eq:sim_res2}\\
    \vec{r}_{y_{d}}  &= \frac{d\vec{y}_d}{dt} - G_d(\vec{x}_d, \vec{x}_a, \vec{y}_d, \vec{y}_a, \vec{\psi})  \label{eq:sim_res3} \\
    \vec{r}_{y_{a}}  &= G_a(\vec{x}_d, \vec{x}_a, \vec{y}_d, \vec{y}_a, \vec{\psi}) \label{eq:sim_res4} 
\end{align}
\end{subequations}
\noindent where $\vec{x}_d$ and $\vec{x}_a$ are respectively the differential and algebraic states of the device and $\vec{y}_{d}$ and $\vec{y}_{a}$ are network differential and network algebraic counterparts. Functions $F_d$, $F_a$, $G_d$ and $G_a$ are the subsystems of equations that define the system of non-linear equations solved at each time step $t$. The terms $\vec{r}_{x_{d}}$, $\vec{r}_{x_{a}}$, $\vec{r}_{y_{d}}$, $\vec{r}_{y_{a}}$ correspond to the residuals of the non-linear system of equations for the dynamic and algebraic states. 

It is important to note that models like \eqref{eq:sim_res1}-\eqref{eq:sim_res4} arise in power systems from the application of \ac{SPT} to \eqref{eq:sim1}-\eqref{eq:sim2} to reduce overall model stiffness as describe in \cite{def_paper}. By selectively zeroing-out some of the differential terms and transforming the model into a system of index-1 \ac{DAE}, the ``slow'' states maintain their differential representation and the ``fast'' states are simplified into the algebraic terms.}

\textbullet~ {\textbf{Mass Matrix Model:} This model is implemented for methods derived from the solution of mechanics problems where the differential terms $\frac{d z_t}{d t}$ are multiplied by constants. It represents system dynamics with the equation $M\frac{d z_t}{d t} = h(z)$. Although mass matrix models may have an arbitrary structure, in \texttt{PSID.jl} the focus is on models where $M$ is a diagonal matrix\footnote{These classes of mass matrix models are also named \emph{Lumped mass matrix} models}. The resulting model is as follows:
\begin{subequations}
\begin{align}
    M_x \frac{d\vec{{x}}}{dt} &= F(\vec{{x}}, \vec{{y}}, \vec{\eta})\label{eq:sim_mm1}\\
    M_y \frac{d\vec{{y}}}{dt} &= G(\vec{{x}}, \vec{{y}}, \vec{\psi})\label{eq:sim_mm2}
\end{align}
\end{subequations}
\noindent where $M_x$ corresponds to the mass matrix for the device states and $M_y$ corresponds to the matrix for the network states. The diagonal elements of $M_x$ are determined by the time constants of the device models and can be zero if the dynamics of the state are not included, a common occurrence in large data sets where filtering dynamics are ignored. For instance, $M_y$ is $0$ in the diagonal when the network is using \ac{QSP} assumptions or it can be populated with the line capacitance if the network needs to be modeled including \ac{EMT} dynamics. Similar to the residual model, \ac{SPT} can be used by setting the entries in the diagonal of $M$ from $\epsilon \to 0$ in \eqref{eq:sim_mm1}-\eqref{eq:sim_mm2} and analyzing the conditions under which those entries reduce model stiffness. 
}

In both models, the modeler can also selectively choose the level of detail for certain portions of the model and account for line dynamics only for those circuits where it is important. The flexibility in the model formulation also enables the use of solvers that employ \ac{BDF} and Backward Euler approaches, as well as collocation algorithms derived from the Radau, Rosenbrock, and Rodas methods. This wide variety of solvers is enabled by \texttt{PSID.jl}'s integration with the \texttt{DifferentialEquations.jl} API \cite{rackauckas2019confederated}, which supports a wide variety of solution methods for both residual and mass-matrix formulations of the simulation model. Integrating the simulation model through a generic solver API allows the \texttt{PSID.jl} framework to be used to formulate large-scale stiff power systems simulation models. This in turn supports the development of novel integration algorithms and improvements in existing methods.

\section{Software design and structure} \label{sec:soft}

\begin{figure}[t]
    \centering
    \includegraphics[width=0.7\columnwidth]{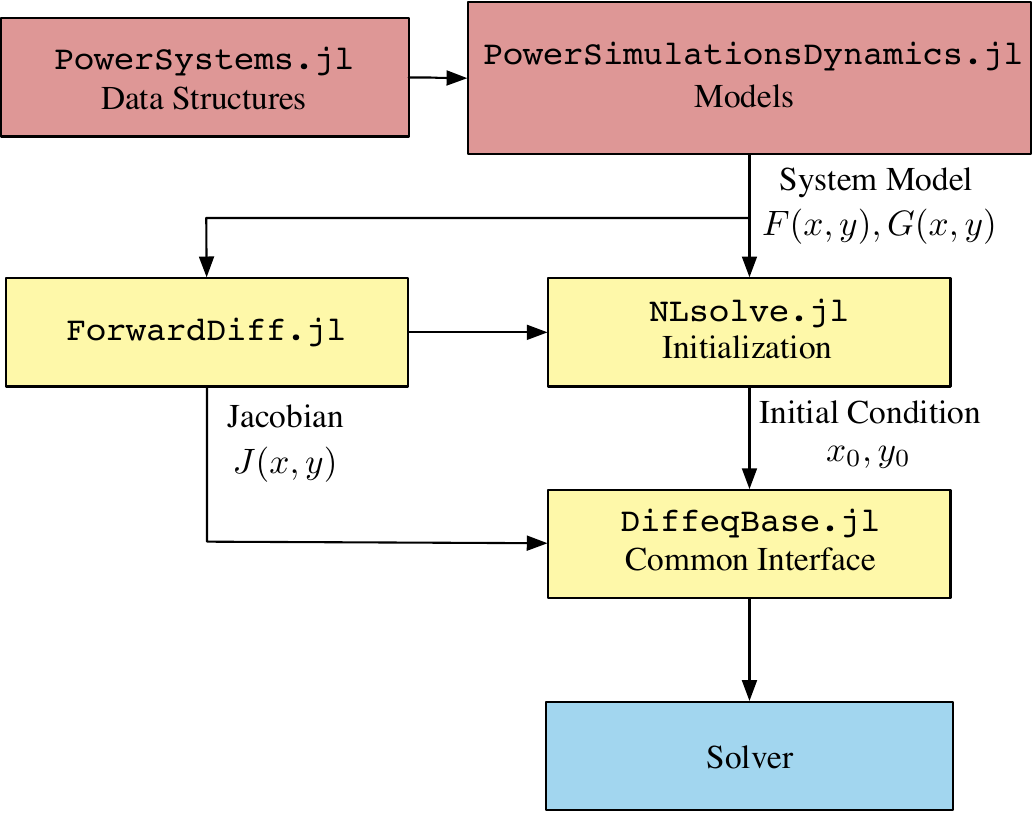}
    \caption{Software dependencies in \texttt{PSID.jl}.}
    \label{fig:InformationFlow}
\end{figure}

\texttt{PSID.jl} is designed to support two primary use cases: 1) package users that want to develop scripts to perform analysis and 2) users that want to develop new device or branch models. The usability objectives require that \texttt{PSID.jl} handles the routines for initialization, Jacobian calculations, interfacing with \texttt{DifferentialEquations.jl}, execution, and post-processing. Another essential advantage of this approach is the implementation of solver callbacks that handle interruption and re-initialization of the integration technique generically.   

Figure \ref{fig:InformationFlow} shows the relationships with the different packages used in \texttt{PSID.jl}. The system model is used to create the Jacobian using \texttt{ForwardDiff.jl} \cite{revels2016forward} and also used to find the initial conditions, in conjunction with \texttt{NLsolve.jl}. Finally, the Jacobian and the initial conditions are passed into the integrator through the common interface (\texttt{DiffeqBase.jl}).

The choice of Julia as the software development environment plays an important role in enabling the desired qualities of power systems simulation, as described in \cite{milano2010power}. Julia is a language that is similar to Python and Matlab, but it offers the performance that one would associate with low-level compiled languages \cite{henriquez2020grid}. Julia supports multiple dispatch and composition, which allows us to design a software and model library that is computationally efficient, yet easy to use and extend through method overloading \cite{methodsjulia}. Julia's multiple dispatch is particularly useful for mathematical modeling since methods can be defined based on abstract data structures. This enables code reuse and easy interfacing with existing models \cite{bezanson2017julia}. \texttt{PSID.jl} evaluates different models by selecting the appropriate method version based on the signature of arguments passed into the function. For instance, Listing \ref{lst::template} shows a prototype of method overload required to develop a model for a \texttt{MyCustomDevice}. The method \texttt{device!()} is dispatched based on the type of the \texttt{device} field, which is defined by the user. 

\begin{listi}[t]
\includegraphics[width=0.98\columnwidth]{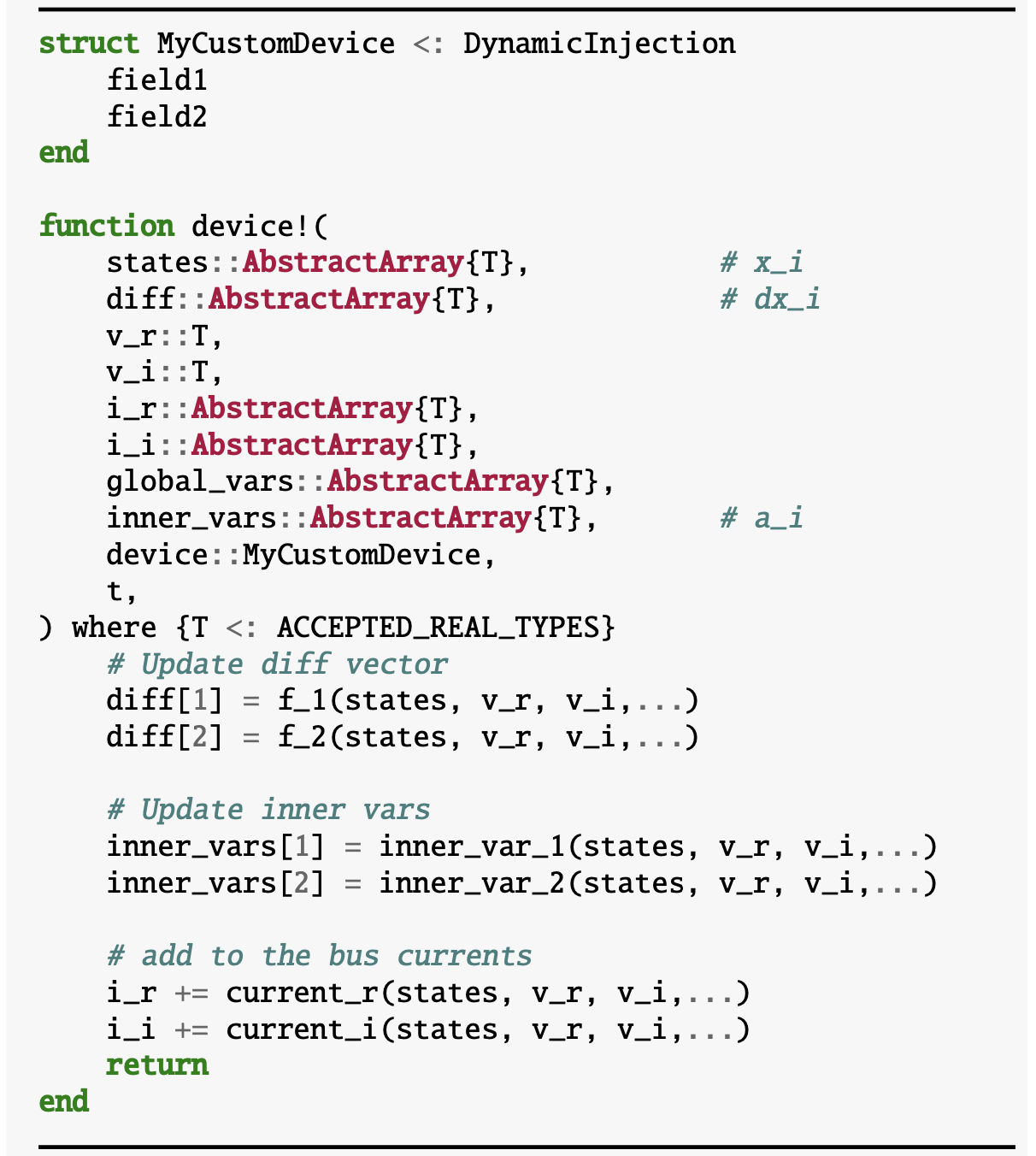}
\captionsetup{font=footnotesize, justification=raggedright,singlelinecheck=false}
\caption{Injection device method prototype} 
 \label{lst::template}
\end{listi}

\subsection{System model implementation}\label{sec:sys}

Equations \eqref{eq:sim_res1}-\eqref{eq:sim_res4} and \eqref{eq:sim_mm1}-\eqref{eq:sim_mm2} are implemented in the software as a Julia structure with pre-allocated vectors to perform in-place-updates of $\frac{d\vec{{x}}}{dt}$, $\frac{d\vec{{y}}}{dt}$, $\vec{{x}}$, $\vec{{y}}$. In addition, a vector $\vec{{a}}$ is used for intermediate inner variables when it is necessary to share information between model components. This design avoids the unnecessary addition of algebraic equations to the state vector, which is a critical consideration because the computational complexity of an implicit method is dominated by the linear solver execution with an upper bound $O(n^3)$ \cite{bojanczyk1984complexity}. The system model also keeps track of global variables to which all models have visibility; one example of this is the system frequency $\omega_\text{sys}$ variable. 

\begin{figure}[t]
    \centering
    \includegraphics[width=0.7\columnwidth]{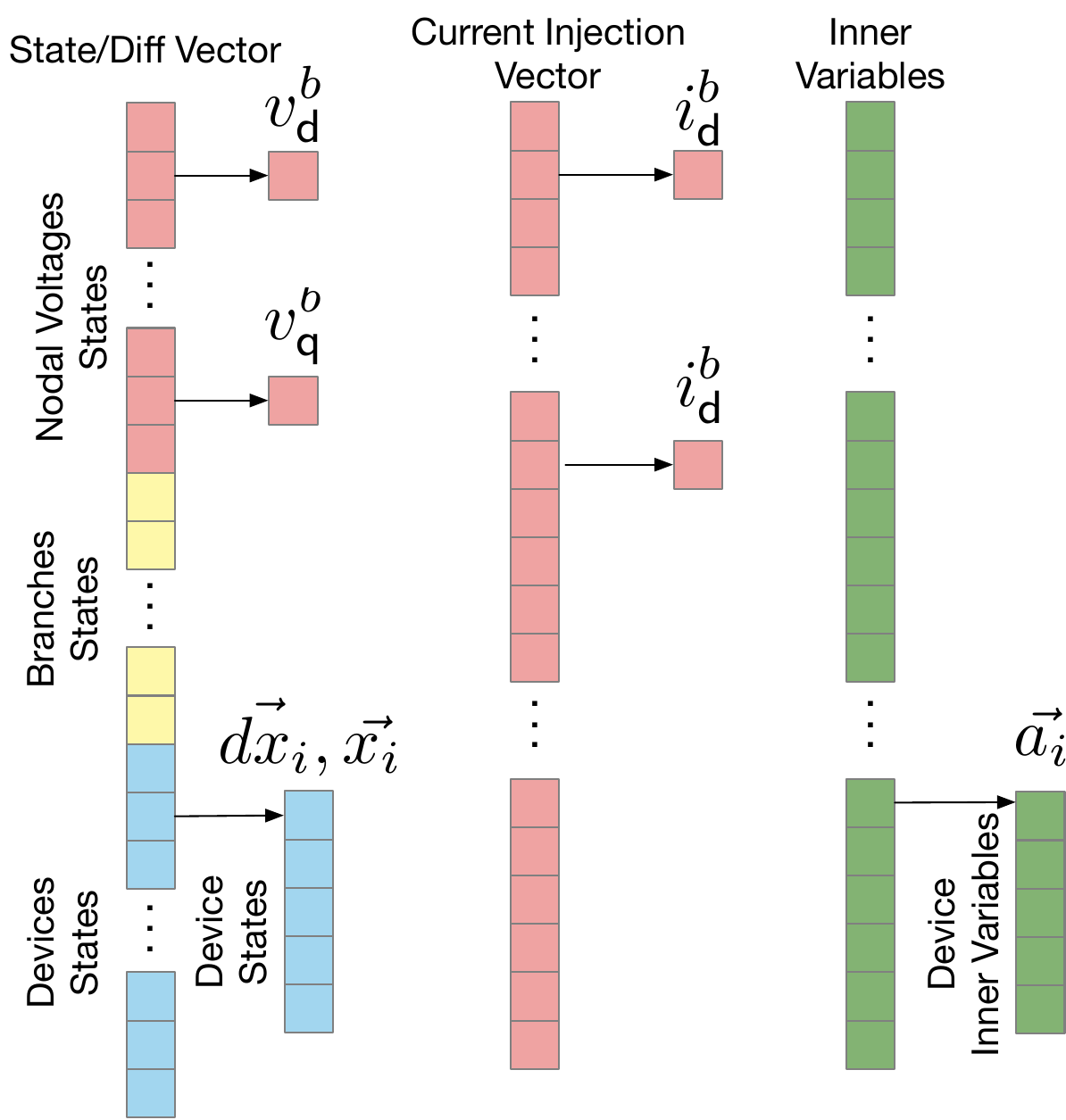}
    \caption{Implementation of the state space indexing.}
    \label{fig:indexing}
\end{figure}

These vectors are stored in a system container to be used as caches. During the simulation build, these caches collect each device together with its states, inner variables, and bus location to generate an index of the cached objects. The index is a tree structure, with the device on the first level and the component on the second level. Figure \ref{fig:indexing} shows the organization of the internal cached vectors in the simulation structure.

\subsection{Jacobian \ac{AD} Approach} \label{sec:autodiff}

Performing Jacobian computations accurately and efficiently is crucial when using implicit solution methods for stiff DAE (Differential Algebraic Equation) systems because the Jacobian matrix is used in the non-linear solving process (usually a Newton Method) and to implement adaptive time-stepping techniques. There are many sparsity techniques available that can minimize the computational cost of Jacobian evaluations, which presents an opportunity to address the challenges of simulating systems with \acp{IBR} at scale. Numerical computation of large scale Jacobians has gained significant attention over the last two decades, especially with the growing demand in applications such as machine learning and optimization. To compute the Jacobian function, we need to derive the matrix function:
\begin{align}
 J(\vec{x}, \vec{y}, \vec{\eta}, \vec{\psi}) = \left[\begin{array}{cc}
   \frac{\partial}{\partial\vec{x}} F(\vec{x},\vec{y}, \vec{\eta}) & \frac{\partial}{\partial\vec{y}} F(\vec{x},\vec{y}, \vec{\eta})\\
   \frac{\partial}{\partial\vec{x}} G(\vec{x},\vec{y}, \vec{\psi}) & \frac{\partial}{\partial\vec{y}} G(\vec{x},\vec{y}, \vec{\psi})\\
   \end{array}\right]
   \label{eq:jacobian}
  \end{align}
The flexibility afforded by \texttt{PSID.jl} makes it challenging to know \textit{a priori} the potential structure of \eqref{eq:jacobian}. Tools that use symbolic layers such as ANDES can exploit symbolic operation libraries to calculate the Jacobian expressions. However, symbolic differentiation isn't always effective since the length of the expressions grows exponentially. 

In \texttt{PSID.jl}, the approach taken is to employ modern \ac{AD} techniques using \texttt{ForwardDiff.jl} \cite{revels2016forward} as the backend. \texttt{ForwardDiff.jl} employs a dual-number approach for the computation of vector Jacobians. Since the Jacobians of dynamic simulation models are square, forward-differentiation is also the most efficient method for \ac{AD} since forward \ac{AD} computes an Jacobian column-wise, whereas reverse \ac{AD} schemes use row-wise computations \cite{margossian2019review}. 

Providing users with performant Jacobian Matrix function evaluations using \ac{AD} techniques that rely on dual number implementation dictates that the caching design specified in Section \ref{sec:sys} needs to be implemented for \texttt{Float} and \texttt{Dual} number types. Since dynamic power systems problems result in sparse Jacobian matrices, it it possible to generate and cache the Jacobian matrix and provide an in-place  calculation of the system model Jacobian for the integration process. Since the Jacobian is obtained from the indexed system model described in Section \ref{sec:soft} it is possible to map the entries of the \ac{AD}-derived Jacobian directly to the states to perform analysis of the reductions necessary for small signal stability. 

\subsection{Injection device modeling}
\begin{figure}[!t]
    \centering
    \includegraphics[width=0.7\columnwidth]{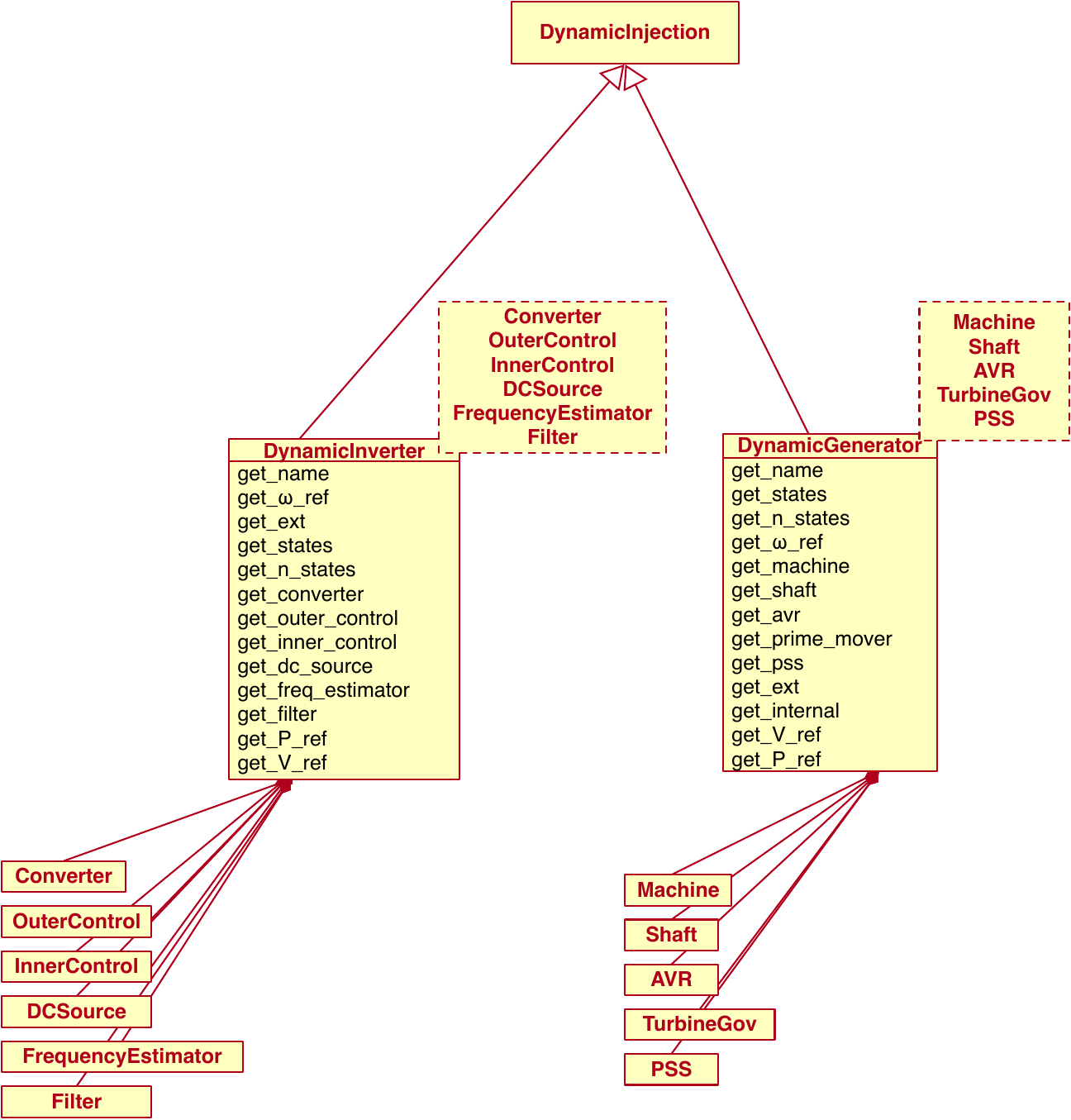}
    \caption{\texttt{DynamicInjection} data structures from \texttt{PowerSystems.jl}.}
    \label{fig:dyn_injection_mm}
\end{figure}
\texttt{PSID.jl} employs the data structures and type hierarchy from \texttt{PowerSystems.jl} \cite{lara2021powersystems} to dispatch the model functions. Generators and inverters are defined as \texttt{DynamicInjection} structures. The data structures use a composition pattern (see Fig. \ref{fig:dyn_injection_mm}), where each injector is defined by individual components. This design enables the interoperability of component definitions within a generic device container allowing for flexible model specifications. As a result, it is possible to implement custom component models and interface them with existing components and models by overloading the \texttt{device!}\footnote{In Julia's diction, functions that mutate arguments have use \texttt{!} character in the name} method. Whenever the function iterates over the vector containing the entire state space, only the relevant portions are accessed by \texttt{device!}. 

Listing \ref{lst::template} shows how within the device modeling layer, the sub-components from the meta-models share local information internal to each device. For instance, in an \ac{IBR} model, the current reference variable $i^{ref}_{olc}$ from the outer loop needs to be shared with the inner loop control, and in a generator model the mechanical torque variable $\tau_m$ is required by the shaft model. This design further modularizes the implementation of the sub-components within an injection device. Intra-device information passing is standardized via port variables that enable efficient information passing between component modeling functions. The arrows in Figs. \ref{fig:inv_metamodel} and Supplemental Information Fig. 1 showcase the port variables and their source-destination relationships for the inverter and generator, respectively.  

Listing \ref{lst::template} also shows a template on how \acp{UDM} can be implemented for new \texttt{DynamicInjection} devices or for new components. A developer has pre-defined API calls that provides systematic access to other device or system variables and use them for multiple control architectures.

\subsubsection{Generator Modeling}
The synchronous machine generator models follow common practices well established in the literature \cite{kundur1994power, milano2010power} on variable sharing and decomposition. The machine model in \texttt{PSID.jl} benefits from the existing standardization in generator models already implemented in most software solutions based on the underlying physics of the generator. Supplemental Information Fig. 1 depicts component levels used in \texttt{PSID.jl} to describe synchronous generators, where each generator is defined by a machine model, an excitation circuit and associated controls (\ac{AVR} and \ac{PSS}), a shaft model, and a prime mover.  

\begin{figure}[t]
    \centering
    \includegraphics[width=\columnwidth]{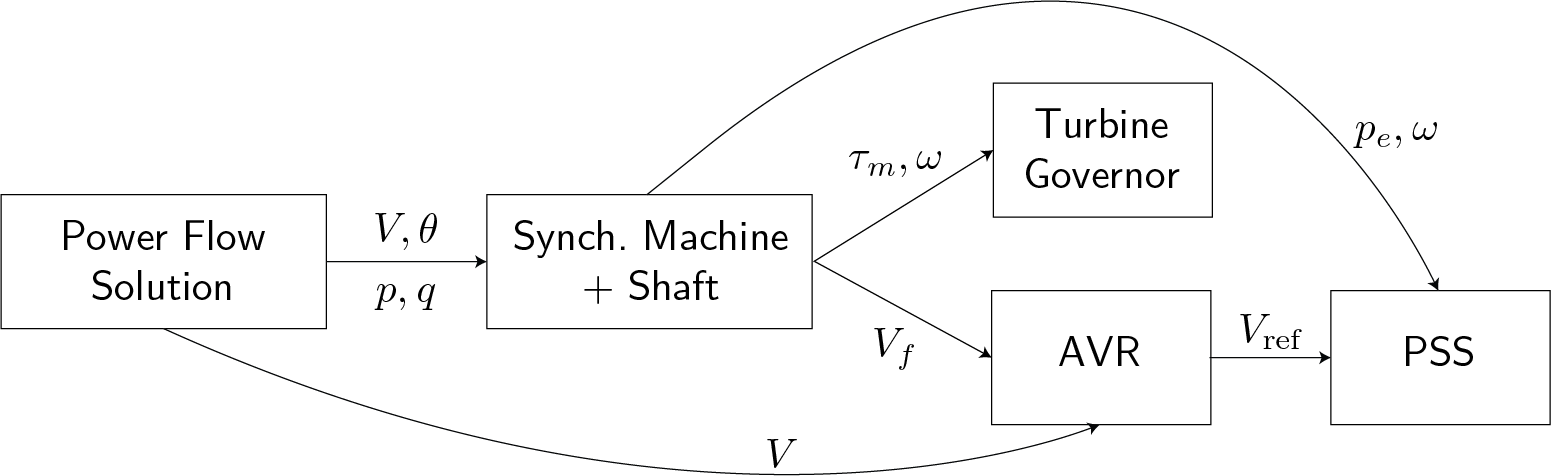}
    \caption{Machine initialization routine.}
    \label{fig:synch_init}
\end{figure}

The synchronous machine metamodel allows the implementation of the initialization routine for generators described in \cite{milano2010power} and shown in Fig. \ref{fig:synch_init}. In this proceedure, we take the power flow solution, initialize the  

\subsubsection{Inverter Modeling}
\begin{figure}[t]
    \centering
    \includegraphics[width=0.7\columnwidth]{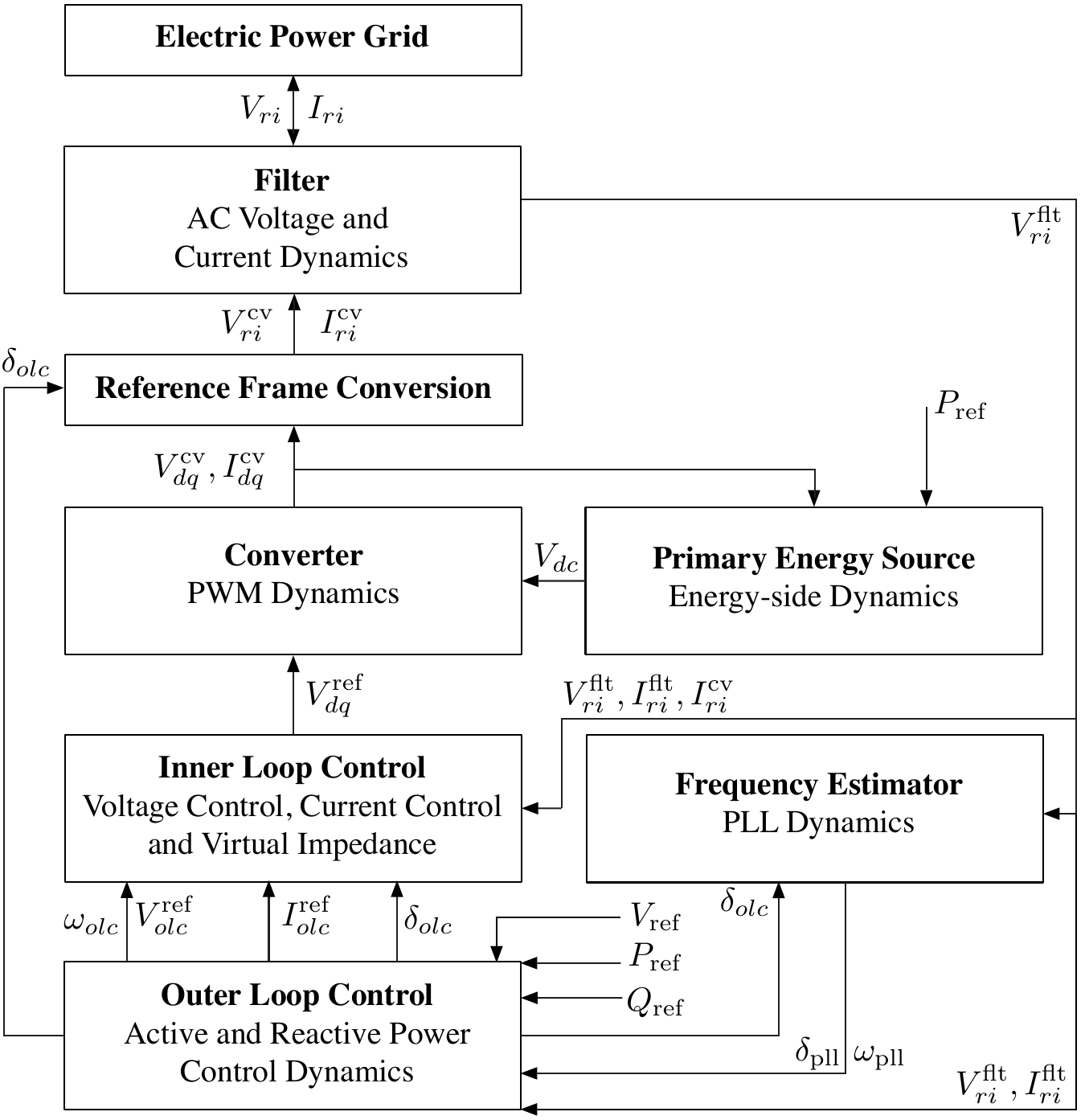}
    \caption{Inverter metamodel.}
    \label{fig:inv_metamodel}
\end{figure}

The proposed \ac{IBR} component decomposition augments the one implemented for \ac{QSP} modeling, which distinguishes between three main modeling blocks: Generator, Electrical, and Plant-level controls. However, the default \ac{QSP} structure cannot integrate frequency measurement modeling and integration with additional control models like \ac{VSM}. Furthermore, the representation of the filter in \ac{QSP} is limited to an algebraic representation due the nature of the simulation tools, which restricts the implementation of the model in \ac{EMT} settings. 
In \texttt{PSID.jl} \ac{IBR} simulation models are structured according to cascaded control architectures, and as a result, there is less standardization of the simulation models. Figure \ref{fig:inv_metamodel} depicts the relationships for the \texttt{DynamicInverter} model as implemented in \texttt{PSID.jl}. The sub-component separation and variable sharing in the inverter is able to support both grid-following industrial models commonly used in \ac{QSP}\footnote{The extension of existing industrial models to the proposed meta model is discussed in detail in \url{https://nrel-sienna.github.io/PowerSimulationsDynamics.jl/stable/generic/}}, generic grid-following models, as well as more advanced control architectures like \ac{VSM} or droop control. Each inverter is defined by a filter, converter model, inner loop control, outer loop control, frequency estimator (typically models employ a \ac{PLL} although one is not required in all applications), and a primary energy source model. 

\begin{figure}[t]
    \centering
    \includegraphics[width=0.7\columnwidth]{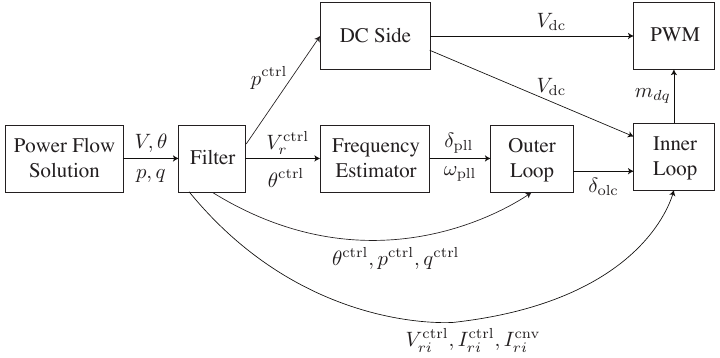}
    \caption{Inverter initialization model.}
    \label{fig:inv_init}
\end{figure}
Since there is no well established initialization method for generic inverter models, \texttt{PSID.jl} employs the sequence described in Fig. \ref{fig:inv_init} to initialize the inverter states. The initialization routine generalizes to any inverter implemented in a compatible fashion with the metamodel. 

\subsection{Load Models}
\texttt{PSID.jl} supports the two major static load models: ZIP and Exponential. Load models are implemented by aggregating the individual loads located at a bus $b$ and using the network \ac{SRF}. Each aggregated load lumps all the currents from the static loads using a rectangular model. All the load models follow the same methods as the injection devices by updating the total current at each bus $b$. For example, the total load currents in a bus with ZIP loads is implemented as follows:
\begin{subequations}
\begin{align}
    i^b_\d &= \frac{1}{|v^b||v_0^b|^2} \sum_{l \in \mathcal{L}^b_i} |v_0^b| \left( p_l v^b_\d + q_l v^b_\q \right) + \nonumber \\
    &\sum_{l \in \mathcal{L}^b_p}  |v_0^b|^2 \left (p_l v^b_\d + q_l v^b_\q \right) + \sum_{l \in \mathcal{L}^b_z}|v^b| \left (p_l v^b_\d + q_l v^b_\q \right) \\ 
    i^b_\q &= \frac{1}{|v^b||v_0^b|^2} \sum_{l \in \mathcal{L}^b_i}  |v_0^b| \left (p_l v^b_\q - q_l v^b_\d \right) + \nonumber \\ 
    &\sum_{l \in \mathcal{L}^b_p} |v_0^b|^2 \left(p_l v^b_\q - q_l v^b_\d \right) + \sum_{l \in \mathcal{L}^b_z}|v^b| \left (p_l v^b_\q - q_l v^b_\d  \right)
\end{align}
\end{subequations}
\noindent where $\mathcal{L}^b_p$, $\mathcal{L}^b_i$, $\mathcal{L}^b_z$ are the sets of constant power, constant current and constant impedance loads at each bus $b$ respectively. $v^b_0$, $p_l$, $q_l$ are the bus voltage and load's power from the power-flow used to estimate the total load currents $i^b_\d$ and $i^b_\q$ at the bus. In addition to the static load models, several models of dynamic loads are available including 5- and 3-state induction machine models \cite{krause2013analysis} and constant power loads compatible with \ac{QSP} and \ac{EMT} studies \cite{henriquezauba2022smallsignal}.

\subsection{Network Circuit Modeling}
\texttt{PSID.jl} implements a lumped parameter $\pi$-circuit of a transmission line in a $\dq$ reference frame. 
This general representation of the network can capture the circuit network dynamics depending on the requirements for the study. It also enables the modeler to selectively choose which specific branches should be modeled including circuit dynamics. The model is implemented by splitting the real and imaginary portions of the network quantities using a \emph{rectangular} representation as follows: 
\begin{align}
    \frac{1}{\Omega_b} \begin{bmatrix}
L & 0 \\
0 & L 
\end{bmatrix} \frac{d}{dt} \begin{bmatrix} \vec{i}^\ell_\d \\ \vec{i}^\ell_\q \end{bmatrix}\! &= \! E_\ell  \begin{bmatrix} \vec{v}_\d \\ \vec{v}_\q \end{bmatrix} - \begin{bmatrix}
R & -L \\
L & R 
\end{bmatrix} \begin{bmatrix} \vec{i}^\ell_\d \\ \vec{i}^\ell_\q \end{bmatrix} \label{eq:volt}\\
   \frac{1}{\Omega_b} \begin{bmatrix}
\frac{1}{B} & 0 \\
0 & \frac{1}{B} 
\end{bmatrix} \frac{d}{dt} \begin{bmatrix} \vec{v}_\d^{b} \\ \vec{v}_\q^b \end{bmatrix}\! &= \! \begin{bmatrix} \vec{i}_\d^b \\ \vec{i}_\q^b \end{bmatrix} \!  - \!  \begin{bmatrix}
\Re(Y_a) & \Im(Y_a) \\
-\Im(Y_a) & \Re(Y_a) 
\end{bmatrix}\! \begin{bmatrix} \vec{v}_\d^b \\ \vec{v}_\q^b \end{bmatrix}  \label{eq:current-inj}
\end{align}
\noindent where \eqref{eq:volt} represents the Kirchoff Voltage Laws across the series element of a branch, and is used to evolve the electric current of branches modeled dynamically. $E_\ell$ is the rectangular incidence matrix that returns the difference between the sending and receiving buses, and $R$ and $L$ are the the series resistance and inductance matrices of each dynamic branch.

Equation \eqref{eq:current-inj} represent Kirchoff Current Laws across the network (for each bus), where $\vec{i}_\d^b$ and $\vec{i}_\q^b$ are the total real ($\d$-axis) and imaginary current ($\q$-axis) injections from devices and dynamic branches at the bus $b$. The blocks $1/B$ on the left-hand side corresponds to the total lump capacitance at the nodes as described in \cite{def_paper}. If the model doesn't include dynamics lines, the left-hand-side of \eqref{eq:current-inj} is zero yielding the equivalent \ac{QSP} network representation $0 = V - \boldsymbol{Y}I$. 

Since the square matrices on the left and right hand side are sparse and have the same pattern in each block, \texttt{PSID.jl} implements a sparse matrix vector multiplication using the \texttt{nzrange} to reduce computational effort. Note that effectively, the matrices on the left hand side of the equations is $M_y$, the mass matrix in equations \eqref{eq:sim_mm2}.

\subsection{Initialization}

Time domain simulation initialization is notoriously difficult, more so on \ac{EMT} simulations \cite{init_paper_wallace} due to the time-variant nature of the model \cite{def_paper}. In order to initiate the simulation, a stable steady-state condition needs to be obtained for equations \eqref{eq:sim1}-\eqref{eq:sim2}. Finding the equilibrium is achieved by setting the derivative terms to zero and solving the system of non-linear equations. However, without a reasonable first guess, finding a solution can lead to initialization failures at a system level.

\begin{figure}[t]
    \centering
    \includegraphics[width=0.5\columnwidth]{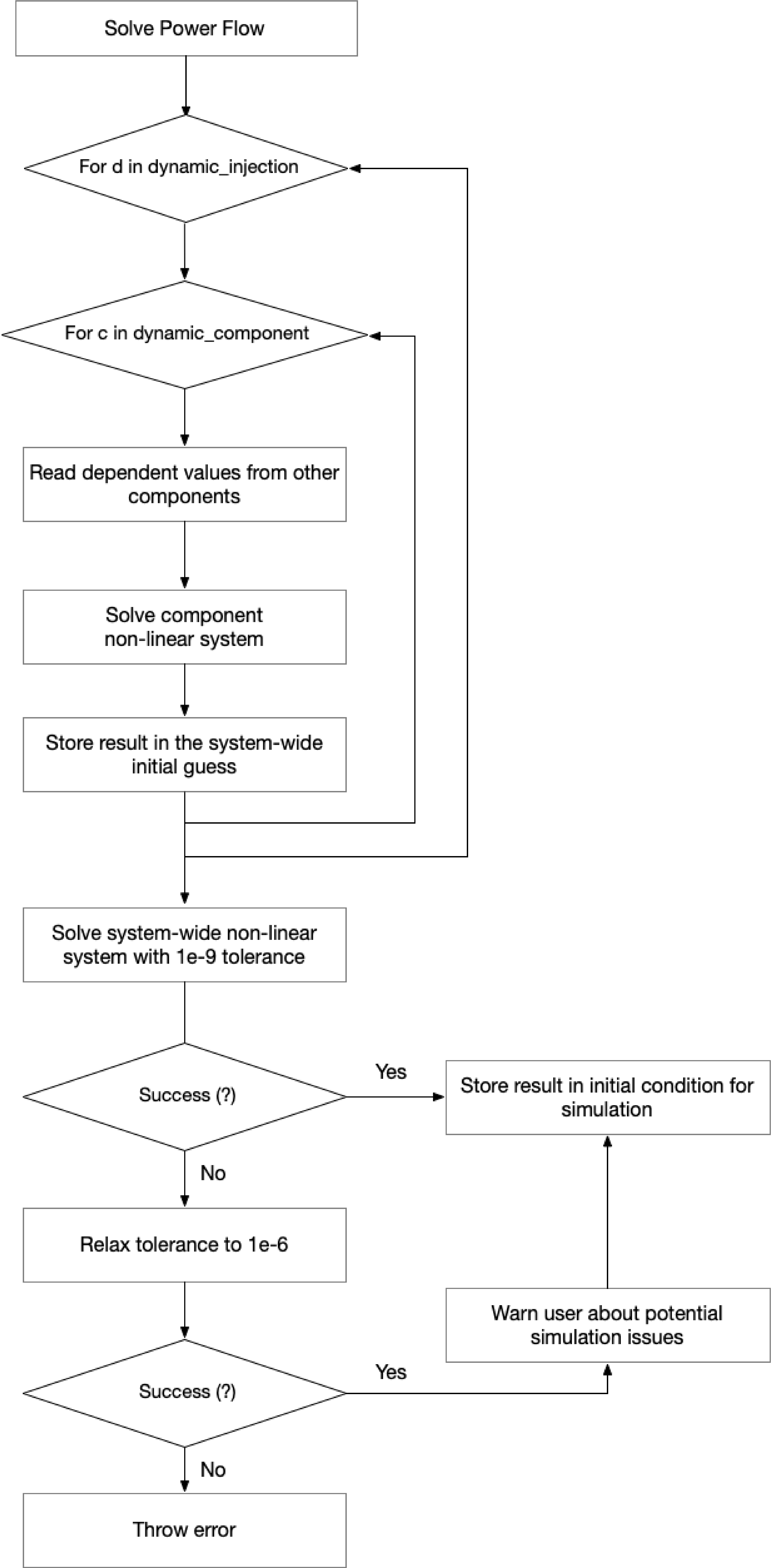}
    \caption{System Level Initialization.}
    \label{fig:sys_init}
\end{figure}

\texttt{PSID.jl} implements an initialization routine illustrated in Figure \ref{fig:sys_init}. This routine ensures that a stable equilibrium exists for every component, device, and the entire system. The initialization sequence for components and devices follows the same procedure as shown in Figures \ref{fig:synch_init} and \ref{fig:inv_init}. This process enables the user and developer to identify the specific states within the components that did not reach equilibrium for debugging and correction purposes.

\subsection{Perturbations}

Perturbations constitute a significant challenge in developing simulation software because the re-initialization procedure can differ depending on the solver. The re-initialization algorithm is different depending on whether the perturbation introduces a change in algebraic or dynamic portions of the model. \texttt{PSID.jl} simplifies the implementation of perturbations (e.g., line faults and step changes) through integrator-agnostic callbacks that considers the appropriate re-initialization technique and avoids the introduction of discontinuities. 

Callbacks enable controlling the solution flow and intermediate initialization without requiring simulation re-starts and other heuristics typically used in power systems dynamic modeling, streamlining the simulation workflow. Each callback defines a function (i.e., the ``affect") that is executed when the simulation reaches a particular condition (e.g, $t = 1.0 $).

Supplement Information Fig. 2 shows the implementation of an impedance multiplication perturbation using integrator-independent callbacks. Modelers can extend the perturbation library by defining custom callback structures, which allows further flexibility in the type of modeling required, spanning from faults to trigger events. 

\section{Small Signal Stability Analysis}

Take the residual formulation \eqref{eq:sim_res1}-\eqref{eq:sim_res4} from dynamic components $\vec{{x}} := (\vec{x}_d^\top, \vec{y}_d^\top)$ and algebraic variables $\vec{{y}} := (\vec{x}_a^\top, \vec{y}_a^\top)$. These are used to define  $S(\cdot) := (F_a(\cdot)^\top, G_a(\cdot)^\top)$ as the vector of differential equations and $R(\cdot) := (F_d(\cdot)^\top, G_d(\cdot)^\top)$ as the vector of algebraic equations of the entire system. Setting the residuals to zero, the resulting non-linear differential algebraic system of equations can be written as:
\begin{align}
\left[\begin{array}{c}
 \frac{d}{dt}\vec{x} \\
  0
  \end{array}\right] = \left[\begin{array}{c}
  S(\vec{x},\vec{y}) \\
  R(\vec{x},\vec{y}) \end{array}\right]
\end{align}
We are interested in the stability around an equilibrium point $\vec{x}_{eq},\vec{y}_{eq}$ that satisfies $d\vec{x}/dt = 0$, or equivalently $S(\vec{x}_{eq}, \vec{y}_{eq}) = 0$, while satisfying $R(\vec{x}_{eq}, \vec{y}_{eq}) = 0$. We first use a first order approximation:
\begin{align}
\left[\begin{array}{c}
 \frac{d}{dt} \Delta \vec{x} \\
  0
  \end{array}\right] = \underbrace{\left[\begin{array}{c}
  S(\vec{x}_{eq},\vec{y}_{eq}) \\
   R(\vec{x}_{eq},\vec{y}_{eq}) \end{array}\right]}_{ = 0}
 + J[\vec{x}_{eq}, \vec{y}_{eq}] \left[\begin{array}{c}
 \Delta \vec{x} \\
  \Delta \vec{y}
  \end{array}\right]
  \end{align}
The Jacobian matrix $J[\vec{x}_{eq}, \vec{y}_{eq}]$ can be split in four blocks depending on the specific variables and functions modeled:
\begin{align}
J[\vec{x}_{eq}, \vec{y}_{eq}] =
\left[\begin{array}{cc}
 S_x & S_y \\
 R_x & R_y \\
  \end{array}\right].
\end{align}
\noindent Assuming that $R_y$ is not singular, we can eliminate the algebraic variables to obtain the reduced Jacobian:
\begin{align}
J_{\text{red}} = S_x - S_y R_y^{-1} R_x ~ \to ~  \frac{d}{dt}\Delta\vec{{x}} = J_{\text{red}} \Delta \vec{x}
\end{align}
that defines our reduced system for the differential variables, from which we can compute its eigenvalues to analyze local stability. 

\section{Simulation Validation and Case Studies} \label{sec:results}

This section presents comparisons of simulation using \texttt{PSID.jl}, the \ac{QSP} simulator PSS\textregistered{}E, and the waveform simulation \ac{EMT} platform PSCAD. We also show the verification results of a large \ac{QSP} case with large penetration of \acp{IBR} and small-signal capabilities using the results as published in ANDES \cite{cui2020hybrid}\footnote{All the simulation code and results is available at \url{https://github.com/NREL-Sienna/PSIDValidation}}. Listing \ref{lst::simulation} shows the \texttt{PSID.jl} code used to run the simulations with a Residual Model and the Sundials solver \texttt{IDA()}.  

\begin{listi}[t]
\includegraphics[width=0.98\columnwidth]{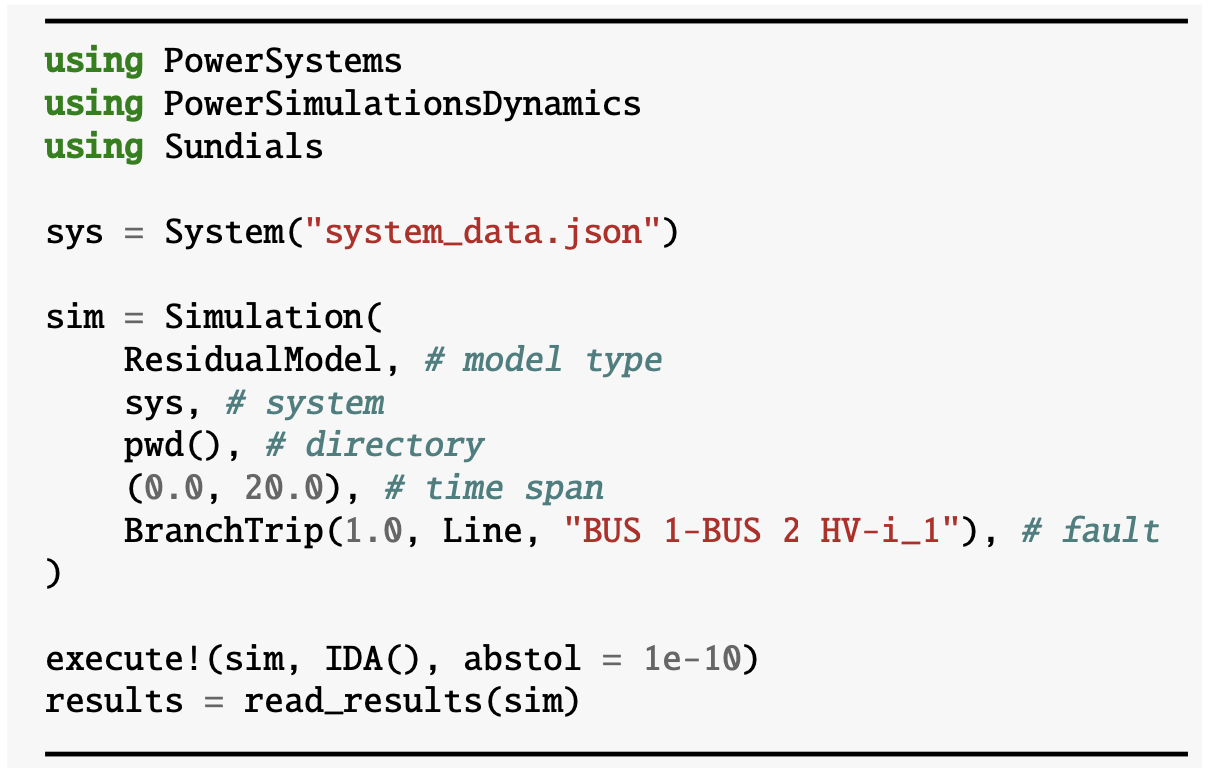}
\captionsetup{font=footnotesize, justification=raggedright,singlelinecheck=false}
\caption{Example of setting up a simulation} 
 \label{lst::simulation}
\end{listi}

\subsection{240-bus WECC case QSP Simulation}

We tested \texttt{PSID.jl} on the 240-bus WECC case for the study of large-scale penetration of \ac{IBR} \cite{yuan2020developing} to assess its modeling capabilities at larger scales employing \ac{QSP}. The system is comprised of the devices described in Table \ref{tab:model_types_240_bus_case} and 329 Lines. The resulting dynamic model has a total of 2420 dynamic states and 506 algebraic states. 

\begin{table}[t]
\caption{Models used in WECC 240-Bus \ac{QSP} phasor validation}
\begin{tabular}{c|ccccc}
Count & Machine & Excitation & Governor  &  PSS   &            \\ \hline
6       & GENROU  & SEXS       & GAST      &  IEEEST &            \\
41       & GENROU  & SEXS       & GAST     &        &            \\
3       & GENROU  & SEXS       & HYGOV      & IEEEST  &            \\
22       & GENROU  & SEXS       & HYGOV      &        &            \\
1      & GENROU  & SEXS       &  TGOV     &   IEEEST     &            \\
36       & GENROU  & SEXS       & TGOV      &        &            \\ \hline \hline
Inverters  & Outer   & Inner      & Converter & Filter & Freq. Est. \\ \hline
121      & REPCA1  & REECB1     & REGCA1    & None   & None       \\
\end{tabular}
\label{tab:model_types_240_bus_case}
\end{table}

We executed 329 line trips and 195 generator trips in the system using PSS\textregistered{}E and \texttt{PSID.jl} and for each contingency, we calculated and compared the Root Mean Square Error (RMSE) for every bus voltage, angle, and generator speed, resulting in a total of 311,780 individual traces to verify the validity of \texttt{PSID.jl} simulation results with respect to a commercial tool for \ac{QSP} modeling PSS\textregistered{}E. We employed the \texttt{Rodas5P()} solver with adaptive time-stepping and a $10^{-9}$ absolute tolerance and PSS\textregistered{}E using a $\Delta t = 0.005$. Table \ref{tab:sim_qsp_errors} shows the results of this exercise; the average error is the average of the max errors over the set of perturbations.  

\begin{table}[t]
\centering
\caption{\ac{QSP} Simulation Error Analysis}
\begin{tabular}{l|cc} \specialrule{.1em}{.05em}{.05em} 
         & Generator Trip   &                                  Line Trip                     \\
        \hline \\[-1.0em] 
Maximum Angle RMSE [deg]        & $4.10 \times 10^{-1}$                  &  $4.50 \times 10^{-1}$ \\
Maximum Voltage RMSE [pu]       & $2.59\times10^{-5}$ &  $2.62\times10^{-5}$   \\
Maximum Speed RMSE [pu]         & $6.57\times10^{-6}$ &  $1.22\times10^{-5}$   \\
Average Angle RMSE [deg]    &  $3.17\times10^{-2}$ & $1.57\times10^{-4}$    \\
Average Voltage RMSE [pu]   & $1.08\times10^{-5}$  & $7.83\times10^{-7}$   \\
Average Speed RMSE [pu]     & $1.60 \times10^{-6}$ &  $5.23\times10^{-7}$   \\ \specialrule{.1em}{.05em}{.05em} 
\end{tabular}
\label{tab:sim_qsp_errors}
\end{table}

\subsection{Balanced \ac{EMT} simulation 144 Bus case} \label{sec:pscad}

In order to verify the simulation capabilities of the $\dq$ -\ac{EMT}, we conducted a waveform simulation on a 144-Bus system using a $\pi$-line. The components used in the simulation are listed in Table \ref{tab:model_types_144_bus_case}. We employ Sauer-Pai machine models \cite{sauer2017power} in order to incorporate stator dynamics. The grid-forming inverters with a droop-control \cite{markovic2021understanding}, and grid-following models use a the current control as detailed in \cite{sajadi2023dynamics}\footnote{The detailed equations of all models can be found in the documentation's component library section https://nrel-sienna.github.io/PowerSimulationsDynamics.jl/stable/}. The inverter model control blocks have been implemented as customized PSCAD components to match the models available in \texttt{PSID.jl}. 

\begin{figure}[t]
    \centering
    \includegraphics[width=0.86\columnwidth]{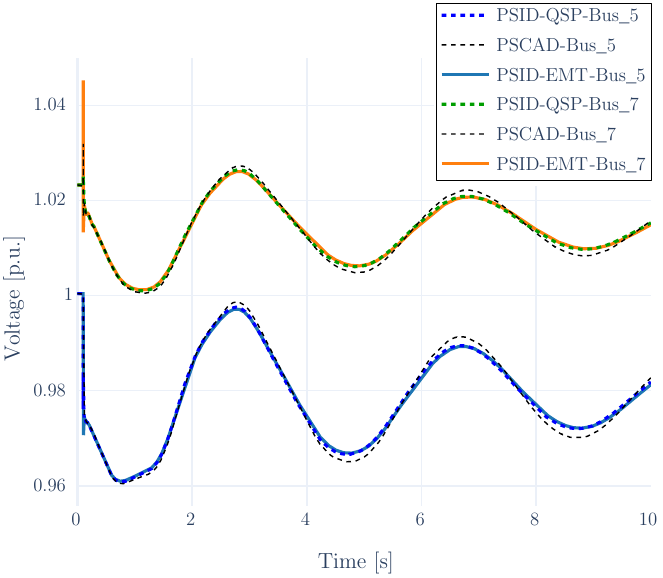}
    \includegraphics[width=0.86\columnwidth]{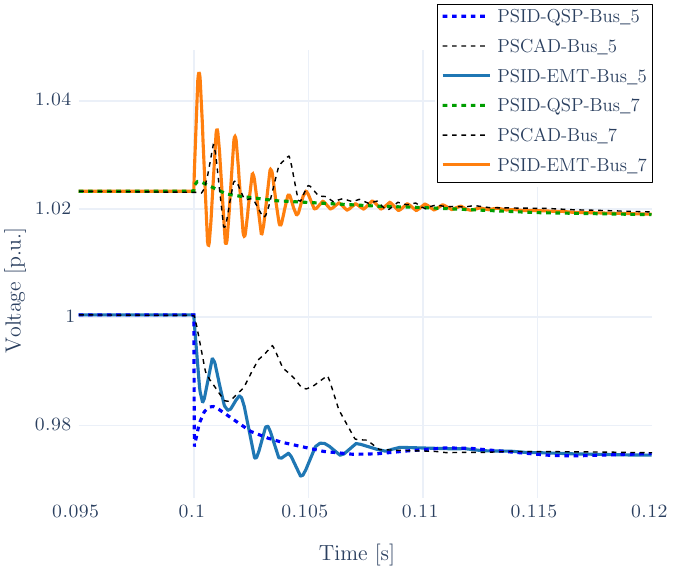}
    \caption{Voltage traces for 144-Bus system line trip.}
    \label{fig:pscad_voltages}
\end{figure}

In this system we conducted a line trip between buses 1 and 2 using a residual model with $\text{abstol}=10^{-9}$ in \texttt{PSID.jl}, and setting 25$\mu$s time step in PSCAD. Additionally, we conducted a simulation specifying the lines to be modeled algebraically as usually done in \ac{QSP} simulations for comparison between the models. Figure \ref{fig:pscad_voltages} showcases the close match between \texttt{PSID.jl} and PSCAD for the voltages of the buses connected to the line trip. At fast timescales, it is difficult to make a direct comparison between the voltage from a time-invariant model and a point-on-wave model. The voltages from PSCAD in Figure \ref{fig:pscad_voltages} are not a state of the simulation, but rather a measured quantity. The PSCAD voltage is measured using the built-in PSCAD digital RMS voltage measurement component, which introduces some delay and imprecision when compared to PSID in the time period immediately following the line trip. 

A more comprehensive error analysis is shown in Table \ref{tab:sim_emt_errors} where we compare the voltage measurements and use the measured active and reactive power at the buses in both simulation environments for a total of 576 trace comparisons. The results show that the RMSE between the traces is well withing numerical precision tolerance. 

\begin{table}[t]
\caption{Models used in 144-Bus \ac{EMT} validation}
\begin{tabular}{c|ccccc}
Gens & Machine & Excitation & Governor  &  PSS   &            \\ \hline
20       & SauerPaiMachine  & SEXS       & TGOV      &  None    &            \\ \hline \hline
IBR  & Outer   & Inner      & Converter & Filter & Freq. Est. \\ \hline
10      & \makecell{Active Droop \\ Reactive Droop}   & \makecell{Voltage \\ Control}     & \makecell{Average \\ Converter}    & \makecell{LCL \\ Filter}   & None       \\
14      & \makecell{Active PI \\ Reactive PI}  & \makecell{Current \\ Control}     & \makecell{Average \\ Converter}    & \makecell{LCL \\ Filter}   & \makecell{Kaura \\ PLL}      \\
\end{tabular}
\label{tab:model_types_144_bus_case}
\end{table}

\begin{table}[t]
\centering
\caption{\ac{EMT} Simulation Error Analysis}
\begin{tabular}{l|cc} \specialrule{.1em}{.05em}{.05em} 
          &                                  Line Trip                     \\
        \hline \\[-1.0em] 
Maximum Active Power RMSE [pu]       &  $4.45 \times 10^{-6}$ \\
Maximum Reactive Power RMSE [pu]       &  $6.85 \times 10^{-6}$ \\
Maximum Voltage RMSE [pu]      &  $3.03\times10^{-6}$   \\
Maximum Bus Frequency RMSE [pu]         &  $1.35\times10^{-7}$   \\
Average Active Power RMSE [pu]       &  $1.89 \times 10^{-6}$ \\
Average Reactive Power RMSE [pu]       &  $2.24 \times 10^{-6}$ \\
Average Voltage RMSE [pu]   & $1.03\times10^{-6}$   \\
Average Bus Frequency RMSE [pu]    &  $1.16\times10^{-7}$   \\ \specialrule{.1em}{.05em}{.05em} 
\end{tabular}
\label{tab:sim_emt_errors}
\end{table}

In terms of simulation time we compared single core execution for both PSCAD and \texttt{PSID}. Although time comparisons are difficult to perform on equal footing due to the methodological differences, the point-of-wave simulation took a total of 86978 ($\approx$ 24 Hr) seconds split in 37269 seconds for initialization and 49709 ($\approx$ 13 Hr) for execution while for \texttt{PSID} the simulation times were 2665 s when employing the pure Julia solver FBDF \cite{rackauckas2017differentialequations} and 69 s when employing Sundial's \texttt{IDA} solver with the sparse linear solver \texttt{KLU} using adaptive time-stepping in each case. The significant difference in computation times showcases the value of formulating the simulation as a time-invariant model for the \ac{EMT} simulation. 

\subsection{Small Signal analysis}
\subsubsection{Validation with ANDES}

\texttt{PSID.jl} \ac{AD} routine for Jacobian evaluation provides the capability to evalate function \eqref{eq:jacobian} at a desired operating point to analyze the small signal stability of a system. We employ a known small-signal unstable system, the 11-bus Kundur system (see Table \ref{tab:eigs_andes}) to verify the eigenvalues calculation in \texttt{PSID.jl}. We compare the eigenvalues and the damping $\zeta$ with outputs from the Python package ANDES \cite{cui2020hybrid} which provides a similar capability to \texttt{PSID.jl}. ANDES eigenvalue routine analyses has already been validated against the commercial tool DSATools SSAT. The RMSE of the eigenvalues between ANDES and \texttt{PSID.jl} is $\frac{1}{N}|| \lambda_\texttt{PSID} - \lambda_\text{ANDES}||_2 \approx 0.2872$. The results confirm that these small signal analyses closely match other for \ac{QSP} models with packages like ANDES. The small differences observed are explained by power-torque approximations performed in shafts and turbine governor models.
\begin{table}[t]
\centering
\caption{Eigenvalue result comparison for 11-bus system}
\begin{tabular}{cccp{-3cm}cc}
\specialrule{.1em}{.05em}{.05em} 
         & \multicolumn{2}{c}{\texttt{PSID.jl}}                      &               & \multicolumn{2}{c}{ANDES}                     \\ \cline{2-3} \cline{5-6} 
         & Eigenvalue & $\zeta [\%]$ &  & Eigenvalue & $\zeta [\%]$ \\ \cline{2-3} \cline{5-6} 
\#1      & $-37.2633+0j$                 & $100$        &               & $-37.2633+0j$                 & $100$        \\
\#2      & $-37.1698+0j$                 & $100$        &               & $-37.1699+0j$                 & $100$        \\
\#3      & $-36.2028+0j$                 & $100$        &               & $-36.2031+0j$                 & $100$        \\[-4pt]
$\vdots$ & $\vdots$                       & $\vdots$     &               & $\vdots$                       & $\vdots$     \\
\#39     & $0.5381\!-\!2.7415j$             & --$19.26$      &               & $0.5537\!-\!2.7459j$             & --$19.77$      \\
\#40     & $0.5581\!+\!2.7415j$             & --$19.26$      &               & $0.5537\!+\!2.7459j$             & --$19.77$      \\ \specialrule{.1em}{.05em}{.05em} 
\end{tabular}
\label{tab:eigs_andes}
\end{table}

\subsubsection{Comparing Small Signal}

\begin{figure}[t]
    \includegraphics[width=3.0in]{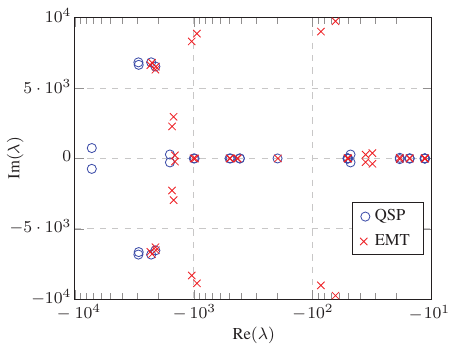}
    \caption{Eigenvalue comparison for balanced EMT simulation.}
    \label{fig:eigs}
\end{figure}

The network modeling flexibility between \ac{QSP} and \ac{EMT} also extends to the small signal analyses allowing a comparison of the eigenvalues between the two network models. Figure \ref{fig:eigs} presents a comparison of the eigenvalues for both formulations in \texttt{PSID.jl} for the system show in Supplemental Information Section 2. 

Our platform offers a valuable feature that assists users in selecting appropriate solvers and model requirements for running simulations. This feature also enables users to analyze the model stiffness, to aid in deciding whether to include EMT dynamics. Furthermore, a thorough review of the eigenvalue properties can help users determine if certain branches can have their differential terms in equations \eqref{eq:volt} and \eqref{eq:current-inj} made zero without introducing significant modeling errors. For instance, in \ref{fig:eigs} employin an algebraic load model will neglect \ac{EMT} dynamics in the 120 Hz range.  

\subsection{Integrator Performance Comparison}

\begin{figure}[t]
    \includegraphics[width=3.2in]{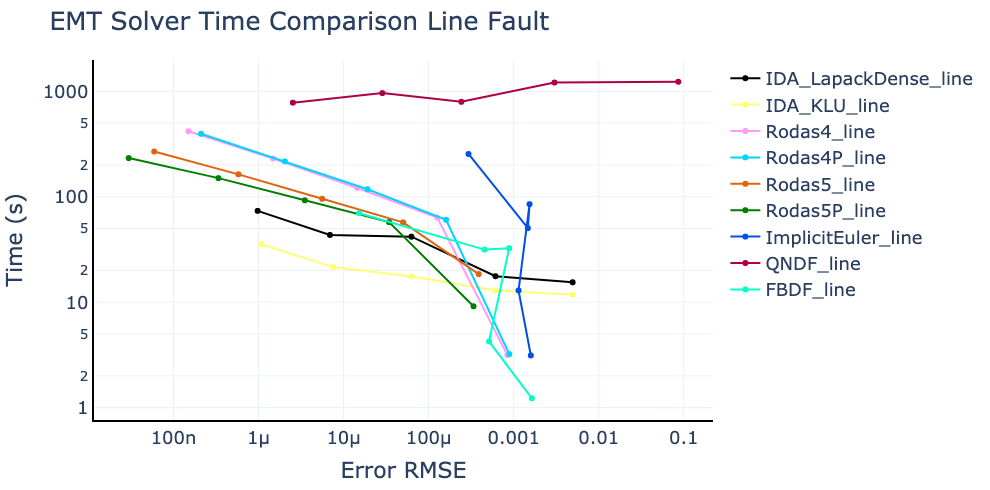}
\includegraphics[width=3.2in]{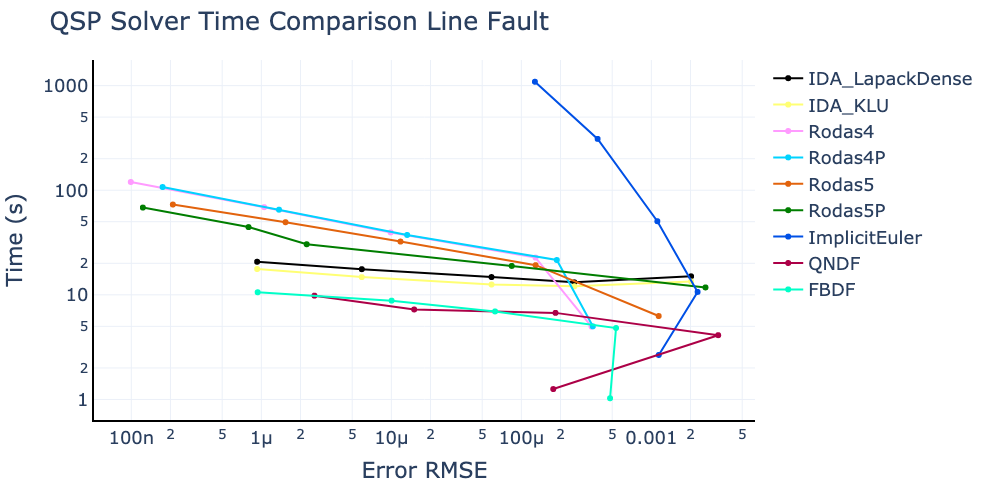}
    \caption{Integrator Work-Precision Diagrams.}
    \label{fig:solver_speed}
\end{figure}

Relying on a single solution method can limit the models that can be simulated in a time-domain simulation. The modular approach between the models and the solvers in \texttt{PSID.jl} also allow to evaluate the performance of the different integration techniques. Figure \ref{fig:solver_speed}\footnote{These results were obtained on a different hardware than the ones in Section V-B} show the precision diagrams of the solvers implemented in \texttt{DifferentialEquations.jl} for both \ac{QSP} and \ac{EMT} formulations of the 144 Bus system used in Section V-B.

The results show that Rosenbrock based methods can find solutions at high precision solutions but with slower solution times. The \ac{EMT} model is significantly stiffer than the QSP case, which changes the ``fastest" solver. The difference in results are a product that certain solver methods will require less number of steps but might have larger solution times due to higher compute costs per integration step.

\section{Conclusion} \label{sec:conclusions}

This paper introduced the open-source toolbox \texttt{PSID.jl}, which focuses on \ac{QSP} and \ac{EMT} time-domain simulations. The proposed software design enables researchers to define a new component's model within the proposed meta-model and quickly explore novel architectures and controls.

Several numerical experiments and benchmarks highlight the capabilities of \texttt{PSID.jl} and the validity of the models included in the library. Case studies show that \texttt{PSID.jl} enables easier assessment of the trade-off between model complexity and computational requirements. The results show that it is possible to replicate waveform \ac{EMT} simulation with a much lower computational cost. 

The ongoing development of  \texttt{PSID.jl} focuses on extending the available models and analytical capabilities. A future key are of development is the inclusion of black-box models and machine learning based surrogates.


%


\bibliographystyle{IEEEtran}
\bibliography{references}

%




\end{document}

%% file: acros2.tex
\begin{acronym}
\acro{AD}{Auto-Differentiation}
\acro{BDF}{Backwards Differentiation Formula}
\acro{EMT}{Electro-magnetic Transient}
\acro{DAE}{Differential Algebraic Equation}
\acro{IBR}{Inverter-based Resource}
\acro{ISO}{Indepent System Operator}
\acro{RMS}{Root Mean Square}
\acro{RCRF}{Rate of Change of the Reference Frame}
\acro{ODE}{Ordinary Differential Equation}
\acro{DE}{Differential Equation}
\acro{PLL}{Phase-locked Loop}
\acro{SPT}{Singular Perturbation Theory}
\acro{SW}{Switching Model}
\acro{AVG}{Average Value Model}
\acro{OEM}{Original Equipment Manufacturer}
\acro{SFA}{Shifted Frequency Analysis}
\acro{PDE}{Partial Differential Equation}
\acro{QSP}{Quasi-Static Phasor}
\acro{MOR}{Model Order Reduction}
\acro{AVR}{Automatic Voltage Regulator}
\acro{PSS}{Power System Stabilizer}
\acro{SRF}{Synchronous Reference Frame}
\acro{UDM}{User Defined Models}
\acro{VSM}{Virtual Synchronous Machine}
\end{acronym}